*Review*

# Simultaneous Control of Bandfilling and Bandwidth in Electric Double-Layer Transistor Based on Organic Mott Insulator κ-(BEDT-TTF)₂Cu[N(CN)₂]Cl

**Yoshitaka Kawasugi** [1,*] **and Hiroshi M. Yamamoto** [2,*]

[1] Department of Physics, Toho University, Funabashi, Chiba 274-8510, Japan
[2] Institute for Molecular Science, National Institutes of Natural Sciences, Okazaki, Aichi 444-8585, Japan
* Correspondence: yoshitaka.kawasugi@sci.toho-u.ac.jp (Y.K.); yhiroshi@ims.ac.jp (H.Y.)

**Abstract:** The physics of quantum many-body systems have been studied using bulk correlated materials, and recently, moiré superlattices formed by atomic bilayers have appeared as a novel platform in which the carrier concentration and the band structures are highly tunable. In this brief review, we introduce an intermediate platform between those systems, namely, a band-filling- and bandwidth-tunable electric double-layer transistor based on a real organic Mott insulator κ-(BEDT-TTF)₂Cu[N(CN)₂]Cl. In the proximity of the bandwidth-control Mott transition at half filling, both electron and hole doping induced superconductivity (with almost identical transition temperatures) in the same sample. The normal state under electric double-layer doping exhibited non-Fermi liquid behaviors as in many correlated materials. The doping levels for the superconductivity and the non-Fermi liquid behaviors were highly doping-asymmetric. Model calculations based on the anisotropic triangular lattice explained many phenomena and the doping asymmetry, implying the importance of the noninteracting band structure (particularly the flat part of the band).

## 1. Introduction

The Mott transition, one of the core subjects in condensed matter physics, allows for the observation of intriguing phenomena, such as high-temperature superconductivity, exotic magnetism, pseudogap, and bad-metal behavior [1]. Although the Hubbard model is thought to include the essential physics of these phenomena, a detailed comparison of the model and real materials is lacking because the pristine Mott state is commonly obscured by a complicated band structure. In addition, the control parameter in a real Mott insulator is usually limited to either bandfilling or bandwidth. Moiré superlattices formed by atomic bilayers have recently emerged as a novel platform for correlated electron systems. Twisted bilayer graphene exhibits superconductivity and correlated insulating states [2], and transition metal dichalcogenide heterobilayers provide a correlation-tunable, Mott-insulating state on the triangular lattice [3]. These artificial systems are a powerful tool to understand the fundamental physics of quantum many-body systems. However, the electronic energy scale of these systems is quite different from that of bulk correlated materials, and an intermediate platform between the artificial and highly tunable moiré superlattices and the bulk correlated materials, such as high-$T_C$ cuprates, is invaluable.

In this brief review, we introduce bandfilling- and bandwidth-control measurements in a transistor device based on an organic antiferromagnetic Mott insulator (Figure 1) [4–6]. We fabricated an electric double layer (EDL) transistor [7], which is a type of field-effect transistor, using an organic Mott insulator. Gate voltages induced extra charges on the Mott insulator surface, which resulted in bandfilling shifts. On the other hand, by bending the EDL transistor, the Mott insulator was subjected to strain, which resulted in bandwidth changes. The (gate voltage)-(strain) phase diagram corresponded to the conceptual phase diagram of the Mott insulator in bandfilling-bandwidth 2D space. We experimentally mapped the insulating, metallic, and superconducting phases in the phase diagram. The experimental phase diagram showed that the superconducting phase surrounded the insulating phase with a particularly doping-asymmetric shape. The asymmetry was partly reproducible by calculations based on the Hubbard model on an anisotropic triangular lattice, implying the importance of the noninteracting band structure. We also showed that the normal states in the doped Mott-insulating state exhibited non-Fermi liquid behaviors, probably due to the partial disappearance of the Fermi surface (FS), similarly to the high-$T_C$ cuprates.

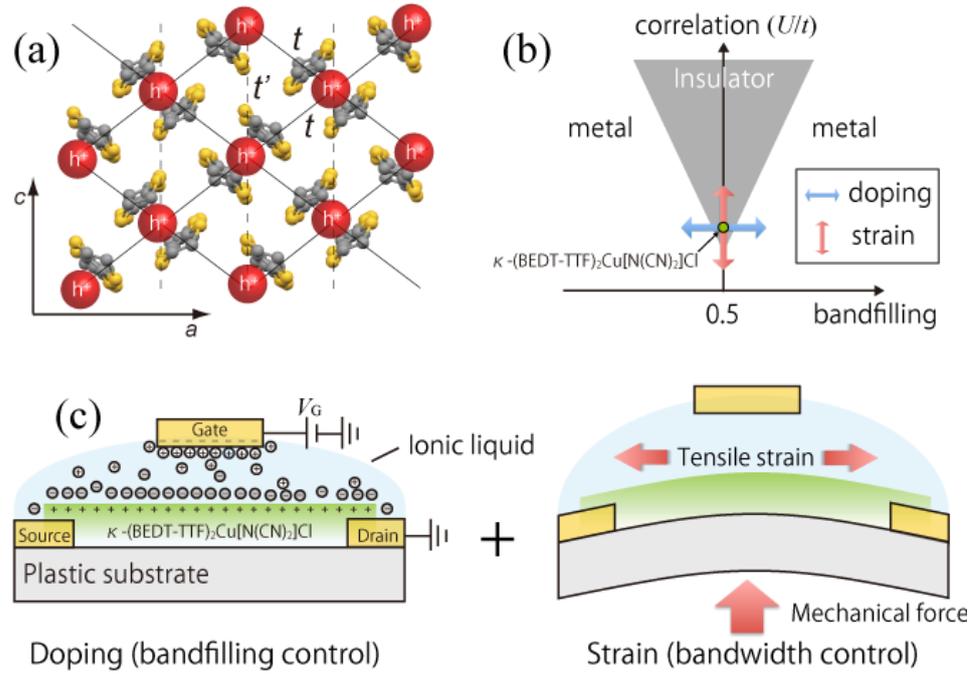

**Figure 1.** (**a**) Conducting BEDT-TTF layer in κ-(BEDT-TTF)$_2$Cu[N(CN)$_2$]Cl. (**b**) Conceptual phase diagram based on the Hubbard model [1]. The vertical axis denotes the strength of the electron correlation. κ-(BEDT-TTF)$_2$Cu[N(CN)$_2$]Cl is located near the tip of the insulating region. (**c**) Schematic side view of the device structure. Doping concentration and bending strain are controlled by EDL gating and substrate bending with a piezo nanopositioner, respectively.

## 2. Materials and Methods

*2.1. Subject Material: Organic Mott Insulator κ-(BEDT-TTF)$_2$Cu[N(CN)$_2$]Cl*

κ-(BEDT-TTF)$_2$Cu[N(CN)$_2$]Cl (BEDT-TTF: bisethylenedithio-tetrathiafulvalene, abbreviated κ-Cl hereinafter) is a quasi-two-dimensional molecular conductor in which the conducting (BEDT-TTF)$_2$$^+$ layer and the insulating Cu[N(CN)$_2$]Cl$^-$ layer are stacked alternately [8]. The unit cell contains four BEDT-TTF molecules forming four energy bands based on the molecular orbital approximation. Two electrons are transferred to the anion layer, resulting in a 3/4 filled system [9]. However, the BEDT-TTF molecules are strongly dimerized, and the upper two energy bands are sufficiently apart from the remaining two bands, resulting in an effective half-filled system. If we regard the two BEDT-TTF dimers in the unit cell (with different orientations) as equivalent, κ-Cl can be modeled as a half-filled, single-band Hubbard model on an anisotropic triangular lattice: $t'/t = -0.44$ [10], where $t$ is the nearest-neighbor hopping, and $t'$ is the next-nearest-neighbor hopping. Similar to the high-$T_C$ cuprates, the sign of $t'/t$ is negative so that the van Hove singularity lies below the Fermi energy (hole-doped side). However, $t'$ exists only for one diagonal of the dimer sites and accordingly, the FS is elliptical (Figure 2).

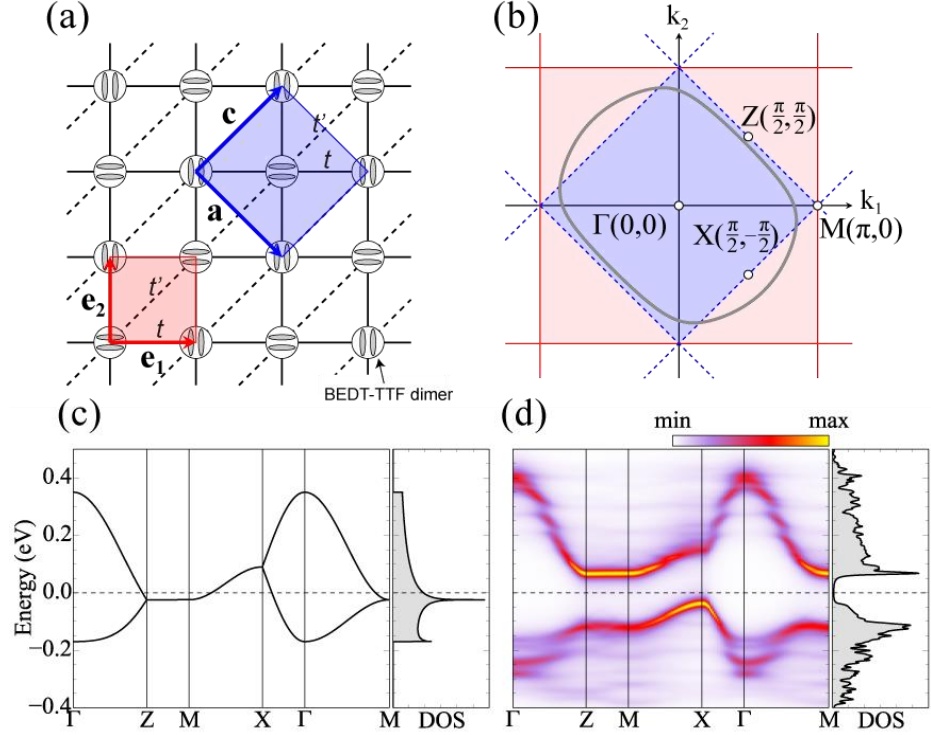

**Figure 2.** Unit cells, Brillouin zones, band structure, and single-particle spectral functions of κ-Cl. Note that the calculations are based on the one-band model. However, the band structure and the spectral functions are shown in the two-site Brillouin zone [blue shaded area in (**b**)] because the adjacent BEDT-TTF dimers are not completely equivalent in the material. Accordingly, the X, Z, and M points in (**c**) and (**d**) correspond to points ($\pi/2$, $-\pi/2$), ($\pi/2$, $\pi/2$), and ($\pi$, 0) in the Brillouin zone of the one-site unit cell. (**a**) Schematic of the anisotropic triangular lattice of κ-Cl. Translational vectors $e_1$ and $e_2$ (*a* and *c*) are represented by red (blue) arrows. The red (blue) shaded region represents the unit cell containing one site (two sites). The ellipses on the sites denote the conducting BEDT-TTF dimers. (**b**) The momentum space for the anisotropic triangular lattice. The Brillouin zones of the one- (two-) site unit cell are represented by the red (blue) shaded region bounded by the red solid (blue-dashed) lines. The solid gray line indicates the FS. (**c**) Noninteracting, tight-binding band structure along highly symmetric momenta and density of states (DOS) of κ-Cl ($t'/t = -0.44$ with $t = 65$ meV). The Fermi level for half filling is set to zero and denoted by the dashed lines. (**d**) Single-particle spectral functions and DOS of κ-Cl at half filling in the antiferromagnetic state at zero temperature, calculated by variational cluster approximation [5]. The Fermi level is denoted by the dashed lines at zero energy. Reproduced with permission from [6].

Because of the narrow bandwidth and the half filling, κ-Cl ($U/t = 5.5$ [10]) is an antiferromagnetic Mott insulator at low temperatures due to the on-site Coulomb repulsion [11]. The material is in close proximity to the bandwidth-control Mott transition. When low hydrostatic pressure is applied (~20 MPa), κ-Cl exhibits the first-order Mott transition to a Fermi liquid/superconductor ($T_C$ ~ 13 K). κ-(BEDT-TTF)$_2$Cu[N(CN)$_2$]Br ($U/t = 5.1$ [10]), which is a derivative with a slightly larger $t$, is also a Fermi liquid/superconductor. The transition has been thoroughly investigated using precise pressure control, such as continuously controllable He gas pressure [12] and chemical pressure by deuterated BEDT-TTF in κ-(BEDT-TTF)$_2$Cu[N(CN)$_2$]Br [13]. As $T_C$ is relatively high for the low Fermi temperatures, the bandwidth-control Mott/superconductor transition in κ-Cl is sometimes regarded as a counterpart to the bandfilling-control Mott-insulator/superconductor transitions in the high-$T_C$ cuprates [14].

κ-Cl has a few hole-doped derivatives. κ-(BEDT-TTF)$_4$Hg$_{2.89}$Br$_8$ has a large $U/t$ (nearly twice that of κ-Cl) but shows metallic conduction and superconductivity because of ~11% hole doping [15]. The transport properties [16,17] are reminiscent of high-$T_C$ cuprates; they show linear-in-temperature resistivity above the Mott–Ioffe–Regel limit (bad-metal behavior) and Hall coefficients inconsistent with the noninteracting FS. Applying

pressure reduces $U/t$, and the temperature dependence of the resistivity approaches that of a Fermi liquid. If the doping concentration is precisely controllable, we would be able to obtain the desired bandwidth–bandfilling phase diagram. However, the doping concentration is fixed, and the doped derivatives are limited (only 11% [15] and 27% [18]).

### 2.2. Experimental Method for Bandfilling Control: EDL Doping

To control the bandfilling of κ-Cl, we employed a doping method based on the EDL transistor [Figure 3a]. The EDL transistor is a type of field-effect transistor in which the gate-insulating film is replaced by a liquid electrolyte such as an ionic liquid. EDL doping enables a higher doping concentration than the typical field-effect doping using a solid gate insulator due to the strong electric fields by the liquid electrolyte. First, we prepared polyethylene terephthalate (PET) substrates and patterned Au electrodes (source, drain, voltage-measuring electrodes, and side-gate electrodes) using photolithography. Next, we synthesized thin single crystals of κ-Cl by electrolysis of a 1,1,2-trichloroethane [10% (*v/v*) ethanol] solution in which BEDT-TTF (20 mg), TPP[N(CN)$_2$] [tetraphenylphosphonium (TPP), 200 mg], CuCl (60 mg), and TPP-Cl (100 mg) were dissolved. We applied 8 μA current overnight and obtained tiny thin crystals of κ-Cl. However, we could not easily remove the thin crystals from the solution because the surface tension of the solution easily broke the crystals. We therefore moved the crystals together with a small amount of the solution by pipetting them into 2-propanol (an inert liquid). Then, using the tip of a hair shaft, we manipulated one crystal and placed it on the substrate in 2-propanol. After the substrate with the κ-Cl single crystal was taken out from 2-propanol and dried, the crystal tightly adhered to the substrate (probably via electrostatic force). The κ-Cl single crystal was shaped into a Hall bar using a pulsed laser beam at the wavelength of 532 nm [Figure 3b]. Lastly, we added a droplet of 1-ethyl-3-methylimidazolium 2-(2-methoxyethoxy) ethyl sulfate ionic liquid on the sample and the Au side-gate electrode and placed a 1.2-μm-thick polyethylene naphthalate (PEN) film on it to make the liquid phase thin. The thinning of the gate electrolyte using the PEN film reduced the thermal stress at low temperatures. We immediately cooled the sample to 220 K (~3 K/min), where the ionic liquid was less reactive. At lower temperatures, the ionic liquid solidified.

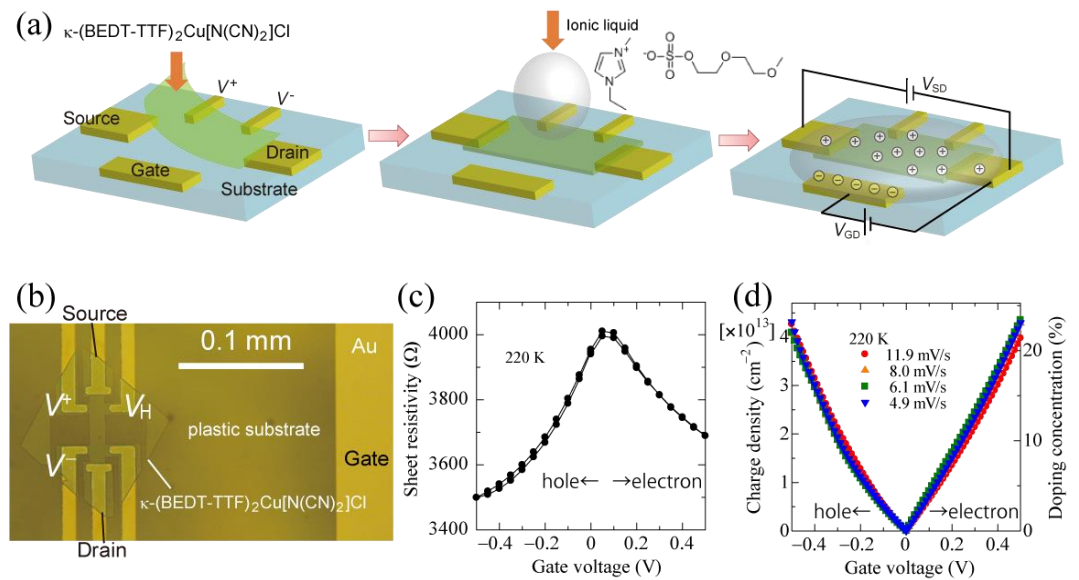

**Figure 3.** (**a**) Schematic view of device fabrication procedure. (**b**) Optical top view of an EDL transistor device. The κ-Cl crystal is laser-shaped into a Hall bar. (**c**) Gate voltage dependence of sheet resistivity at 220 K. (**d**) Gate-voltage dependence of accumulated charge density and doping concentration.

We controlled the doping concentration of the κ-Cl crystal surface by varying the gate voltage, $V_G$, at 220 K. Both the positive and negative gate voltages, corresponding to the electron and hole doping, reduced the sample resistance, implying the deviation of the bandfilling from 1/2 [Figure 3c]. According to the charge displacement current measurements [19], the doping concentration reached approximately ±20% at $V_G$ of ±0.5 V [Figure 3d]. Increasing the gate voltage ($|V_G| > 0.7$ V) led to the irreversible increase of resistance, indicating sample degradation due to chemical reactions. The choice of ionic liquid was important; the crystal immediately disappeared when we employed ionic liquids that were too reactive or too good for the solubilization of κ-Cl. Diethylmethyl(2-methoxyethyl)ammonium bis(trifluoromethylsulfonyl)imide [DEME-TFSI], a typical ionic liquid for EDL doping, was suitable for electron doping but not for hole doping at low temperatures. At the moment, 1-ethyl-3-methylimidazolium 2-(2-methoxyethoxy) ethyl sulfate is the best choice. We focused on the doping effect on this ionic liquid.

*2.3. Experimental Method for Bandwidth Control: Uniaxial Bending Strain via Substrate*

We usually control the bandwidth of a molecular conductor by applying hydrostatic pressure using a pressure medium oil and a pressure cell. However, because the heterogeneous device structure was unsuitable for hydrostatic pressure application, we adopted the strain effect caused by substrate bending, as shown in Figure 4. This method required no liquid pressure medium (the ionic liquid is already on the crystal) and enabled precise strain control by fine tuning the piezo nanopositioner that bent the substrate (Figure 4). Assuming that the bent substrate is an arc of a circle (angle: $2\theta$, curvature radius: $r$), strain $S$ is estimated as

$$S = \frac{2\theta(r + d/2) - 2\theta r}{2\theta r} = \frac{\theta d}{l}$$

where $d$ and $l$ are the thickness and the length of the substrate, respectively. The relationship between the sides of the shaded triangle in Figure 4 gives

$$r \sin\theta \sim r\theta = \sqrt{r^2 - (r-x)^2}, \therefore \theta = \frac{4lx}{l^2 + 4x^2}$$

using the small angle approximation, where $x$ is the displacement of the piezo nanopositioner. As a result,

$$S = 4dx/(l^2 + 4x^2).$$

We employed PET substrates with $d = 177$ μm and $l = 12$ mm, and $x$ was up to 2.5 mm so that the typical value of $S$ was ~1 %. Note that the strain in this experimental setup was tensile and uniaxial. As the strain generated by bending was tensile, we employed a PET substrate with a large thermal expansion coefficient to start the bandwidth scanning from the superconducting region. Biaxial compression of the κ-Cl crystal by the substrate at low temperatures resulted in the superconducting state without bending strain. Therefore, we applied the bending tensile strain to enhance $U/t$ and induce the bandwidth-control Mott transition from the metallic/superconducting side to the insulating side. The strain effect should be dependent on the strain direction. However, we leave the detailed direction dependence to future work because the strain-induced Mott transition could be observed regardless of the strain direction at the moment.

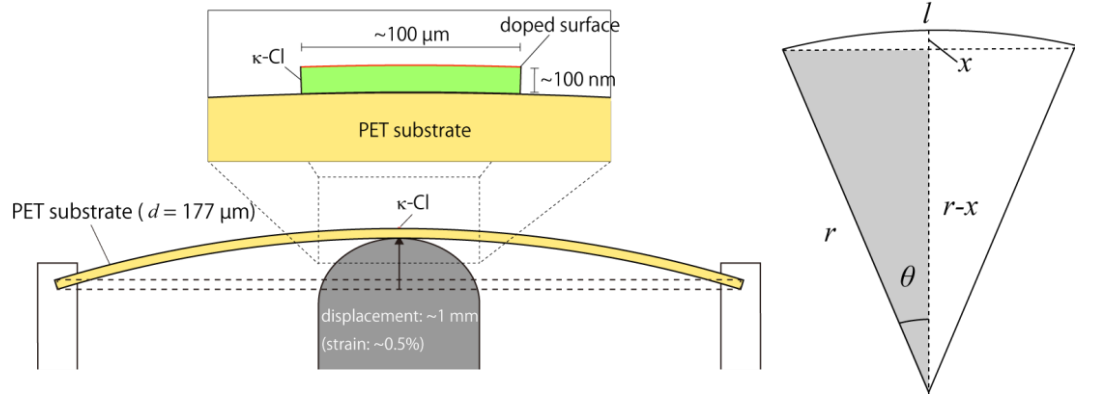

**Figure 4.** Schematic illustration for the application of uniaxial tensile strain to the κ-Cl sample.

## 3. Results

First, we introduced the superconducting phase transitions around the tip of the Mott-insulating state in the bandwidth–bandfilling phase diagram in Section 3.1 [4]. Then, we showed the transport properties under EDL doping at a large $U/t$ in Section 3.2 [5,6] (we did not apply the bending strain shown in Section 2.3 here).

### 3.1. Superconducting Phase around the Mott-Insulating Phase

#### 3.1.1. Strain Effect without Gate Voltage

First, we showed the strain effect without the gate voltage (Figure 5). As mentioned in the Methods section, κ-Cl became a superconductor without bending strain owing to the thermal contraction of the substrate. $T_C$ (~12 K) was similar to that of the bulk κ-Cl crystal under low hydrostatic pressure. The superconducting state disappeared upon applying the uniaxial tensile strain, $S$, and the insulating state appeared at low temperatures. Despite the uniaxiality, the temperature dependence of the resistivity was qualitatively similar to that in the bulk κ-Cl crystal under hydrostatic pressure [Figure 5c]. The slope of the metallic/insulating phase in the phase diagram implied that the insulating state at the lowest temperatures had low entropy and was the antiferromagnetic Mott-insulating state. Thus, we could control the bandwidth (and consequently $U/t$) of the sample across the bandwidth-control Mott transition at half filling.

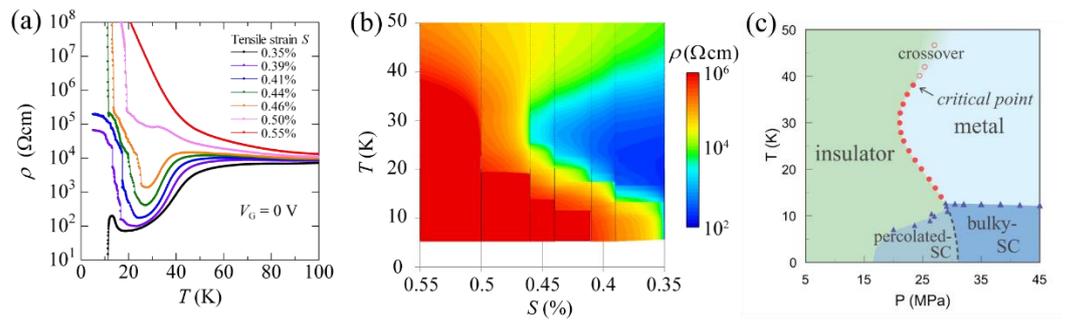

**Figure 5.** (**a**) Resistivity vs. temperature plots under different tensile strains at gate voltage $V_G = 0$ V, and (**b**) contour plots of the resistivity data. (**c**) Pressure-temperature phase diagram of bulk κ-Cl. Reproduced with permission from [12].

#### 3.1.2. Doping Effect at Fixed Strain

Next, we fixed the uniaxial tensile strain at the very tip of the insulating state in the bandwidth–bandfilling phase diagram ($S = 0.41\%$) and applied gate voltages (with warming of the sample to 220 K, changing $V_G$, and cooling again). Both electron and hole doping reduced the resistivity and induced the superconducting state, as shown in Figure 6. $T_C$ was similar (~12 K) among the electron-doped and hole-doped states (and the undoped metallic state). However, the doping effect was highly asymmetric against the

polarity of $V_G$. By hole doping, the resistivity monotonically decreased, and superconductivity emerged for $V_G \leq -0.3$ V [approximately 10% hole doping according to Figure 3d]. On the other hand, the resistivity abruptly dropped, and a superconducting state emerged with low electron doping (+0.14 V $\leq V_G \leq$ +0.22 V, approximately 4~7% electron doping). At the phase boundary, the resistivity discretely fluctuated [5]. After further electron doping, the resistivity increased again, and superconductivity disappeared. Interestingly, the normal-state resistivity at $T > T_C$ also decreased first and increased thereafter with electron doping.

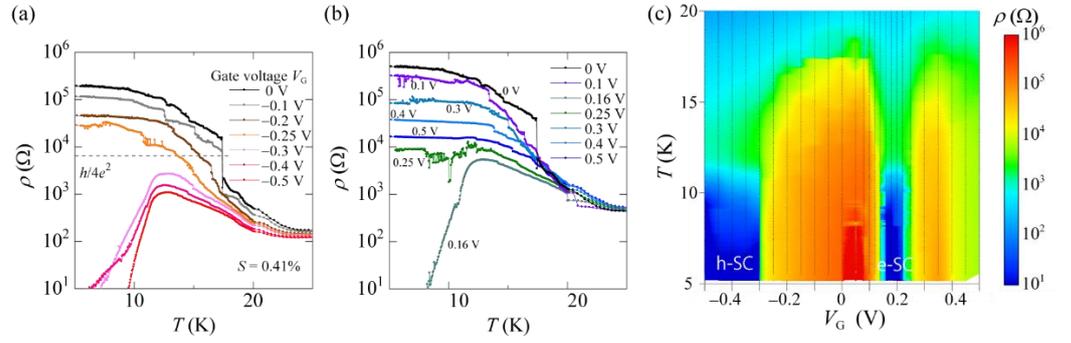

**Figure 6.** Sheet resistivity vs. temperature plots under (**a**) hole doping and (**b**) electron doping at tensile strain $S$ = 0.41%. The dashed line indicates pair quantum resistance $h/4e^2$. (**c**) Contour plots of the data in (**a**) and (**b**). h-SC and e-SC denote hole-doped superconductor and electron-doped superconductor, respectively.

3.1.3. Gate Voltage vs. Strain Phase Diagram

After obtaining the resistivity vs. gate voltage data at the fixed strain, we slightly increased the strain and repeated the same cycles, as shown in Figure 7a. These measurements resulted in the gate voltage vs. strain phase diagram at low temperatures, which corresponded (although not proportionally) to the bandfilling–bandwidth phase diagram, as shown in Figure 7b. The insulating phase was triangular on the hole-doped side (left), similar to the conceptual phase diagram of a Mott insulator, and the superconducting phase surrounded the insulating phase. On the other hand, the superconducting phase appeared to "penetrate" into the insulating phase on the electron-doped side (right).

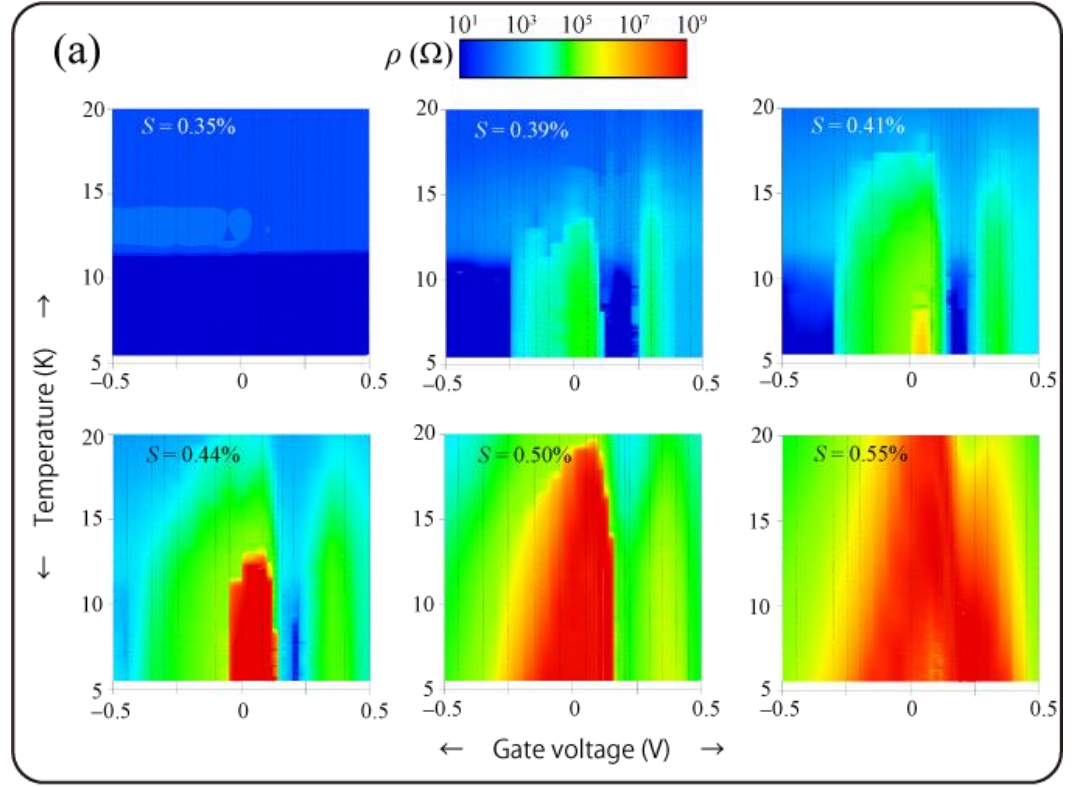

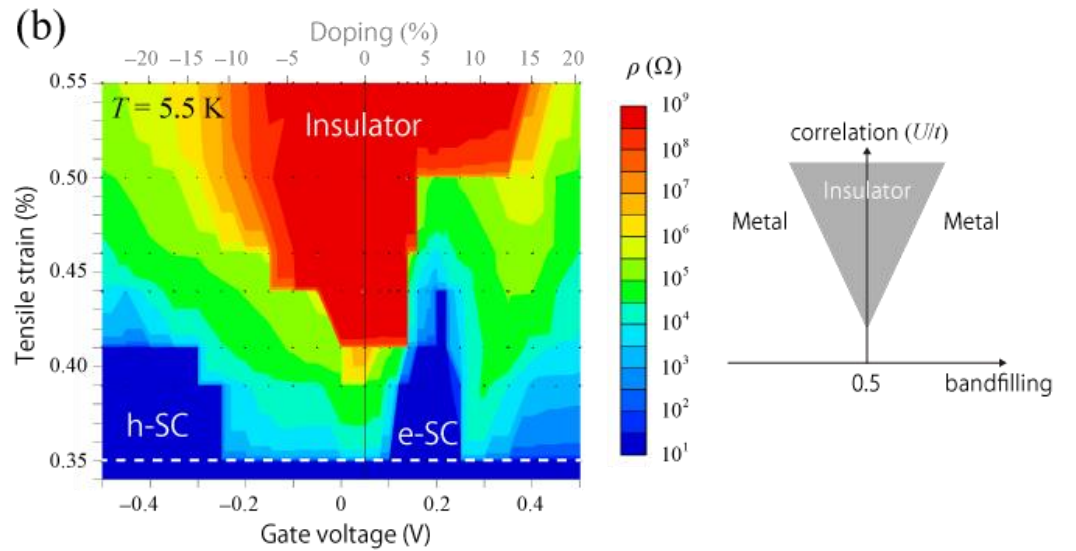

**Figure 7.** (**a**) Contour plots of sheet resistivity, $\rho$, under tensile strains, $S$, of 0.35%, 0.39%, 0.41%, 0.44%, 0.50%, and 0.55% as a function of temperature and gate voltage. (**b**) Contour plots of sheet resistivity, $\rho$, at 5.5 K as a function of gate voltage and tensile strain (left) and the corresponding conceptual phase diagram (right). Black dots in all figures indicate the data points where the sheet resistivity was measured. The doping concentration estimated from the average density of charge accumulated in the charge displacement current measurement [Figure 3d] is shown for reference on the upper horizontal axis in (**b**).

As shown in Figure 8, the doping asymmetry was qualitatively reproduced by variational cluster approximation (VCA) calculations of the antiferromagnetic and superconducting order parameters in a Hubbard model on an anisotropic triangular lattice. At low $U/t$, the superconducting (antiferromagnetic) order parameter more drastically increased (decreased) by electron doping than hole doping [Figure 8a]. The doping dependence of the chemical potential was nonmonotonic only on the electron-doped side [Figure 8b], implying the possibility of a phase separation between the Mott-insulating and

superconducting phases [Figure 8c]. The nonmonotonic behavior of the chemical potential seemed to have originated from the flat part at the bottom of the upper Hubbard band (along the Z–M axis), which was originally located below the Fermi energy in the noninteracting energy band. The flat part of the energy band was caused by the absence of $t'$ along the crystallographic $a$-axis, namely, by the nature of the triangular lattice.

Notice that the experimental results are not understood uniquely within the half-filled band scenario. Calculations on a more detailed quarter-filled band model for the κ-BEDT-TTF salts also predict the doping-induced superconductivity, where the doping polarity alters the pairing symmetry (electron doping: extended $s + d_{x^2-y^2}$, hole doping: $d_{xy}$) [20]. In addition, many quantum Monte Carlo calculations on the Hubbard model indicate the absence of superconductivity at near half filling [21,22], while superconductivity is predicted near quarter filling [23].

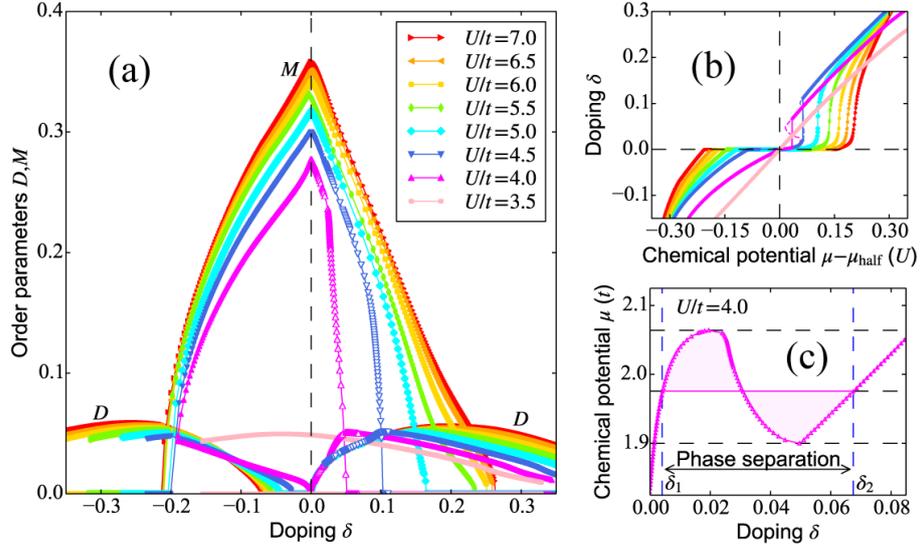

**Figure 8.** VCA calculations. (**a**) Antiferromagnetic and $dx^2 - dy^2$ superconducting order parameters, $M$ and $D$, respectively, vs. doping concentration, $\delta$, for several values of $U/t$. $M$ and $D$ for metastable and unstable solutions (empty symbols) at $U/t$ = 4 and 4.5 under electron doping (corresponding to positive d) are also shown. (**b**) Doping concentration, $\delta$, vs. chemical potential, $\mu$, relative to that at half filling ($\mu_{\text{half}}$) for several values of $U/t$. The results for metastable and unstable solutions at $U/t = 4$ and 4.5 are indicated by dashed lines, and the results obtained by the Maxwell construction are denoted by solid vertical lines. The results imply the presence of phase separation and a first-order phase transition. It is noteworthy that there is a steep (nearly vertical) increase in $\delta$ with increasing $\mu$ for larger values of $U/t$ under electron doping, suggesting a strong tendency toward phase separation. (**c**) Chemical potential, $\mu$, vs. doping concentration, $\delta$, for $U/t = 4$. $\delta_1$ and $\delta_2$ are the doping concentrations of the two extreme states in the phase separation. All results in (a) to (c) were calculated using VCA for the single-band Hubbard model on an anisotropic triangular lattice ($t'/t = -0.44$) with a $4 \times 3$ cluster.

*3.2. Non-Fermi Liquid Behaviors in the Normal State under Doping*

Non-Fermi liquid behaviors, such as the metallic-like resistivity above the Mott–Ioffe–Regel limit and the Hall coefficient inconsistent with the volume of the FS, are ubiquitous features of the normal state of many strongly correlated materials. Here we show that our Mott EDL transistor also exhibited such behaviors.

Due to the principle of the EDL transistor, only the sample surface was doped. However, the nondoped region of the sample was also conductive at high temperatures in κ-Cl. Therefore, to discuss the non-Fermi liquid behaviors at high temperatures, we extracted surface resistivity, $\rho_s$, and surface Hall coefficient, $R_{Hs}$. Assuming a simple summation of the conductivity tensors of two parallel layers (surface monolayer and remaining bulk layers), we derived $\rho_s$ and $R_{Hs}$ from

$$\rho_s = \frac{L}{W}\left(\frac{1}{\rho_{\text{measured}}} - \frac{1}{\rho_{0V}} \times \frac{N-1}{N}\right)^{-1}$$

$$R_{Hs} = \rho_s^2 \left(\frac{R_{H\,\text{measured}}^2}{\rho_{\text{measured}}^2} - \frac{R_{H\,0V}^2}{\rho_{0V}^2} \times \frac{N-1}{N}\right)$$

where $L$, $W$, and $N$ denote the length, width, and number of conducting layers, respectively. Suffixes "measured" and "0 V" stand for the actual measured (combined) values and the values at 0 V (nondoped values), respectively. A sample with a larger $U/t$ (using the PEN substrate that had less thermal contraction) than the previous superconducting sample was measured, and uniaxial tensile strain was not applied.

### 3.2.1. Temperature Dependence of the Resistivity

Figure 9a shows the temperature dependence of the surface resistivity, $\rho_s$, under electron doping. Without the gate voltage, the system was insulating at all measured temperatures (2–200 K). Upon low electron doping ($V_G \sim 0.1\,V$, >3% electron doping), metallic-like conduction ($d\rho_s/dT > 0$) above the Mott–Ioffe–Regel limit, $\rho_{\text{MIR}}$, ($\sim h/e^2$, assuming a two-dimensional isotropic FS) appeared at high temperatures even though the system remained insulating at the lowest temperatures. Although the resistivity was not linear-in-temperature (between linear and quadratic) in this temperature range, this was a bad-metal behavior in the sense that the mean free path of carriers was shorter than the site distance. At $V_G = 0.34\,V$, the resistivity below 50 K also exhibited an insulator-metal crossover across $\rho_{\text{MIR}}$. For $V_G > 0.5\,V$, the temperature dependence of $\rho_s$ approached a Fermi liquid (quadratic in temperature) below 20 K.

Hole doping also induced the bad-metal behavior at high temperatures, as shown in Figure 9b. Although an accurate estimation of the power-law exponent was difficult, the temperature dependence of $\rho_s$ appeared more linear in temperature than in the case of electron doping. The temperature dependence was consistent with the linear-in-temperature resistivity in the hole-doped compound, κ-(BEDT-TTF)$_4$Hg$_{2.89}$Br$_8$, at high temperatures [16,17]. However, we could not observe metallic conduction or the Fermi liquid behavior at low temperatures down to $V_G$ of $-0.6\,V$. Thus, at high temperatures, the bad-metal behavior emerged in a wide doping range except at $V_G = 0\,V$, whereas the Fermi liquid state at low temperatures appeared only under high electron doping.

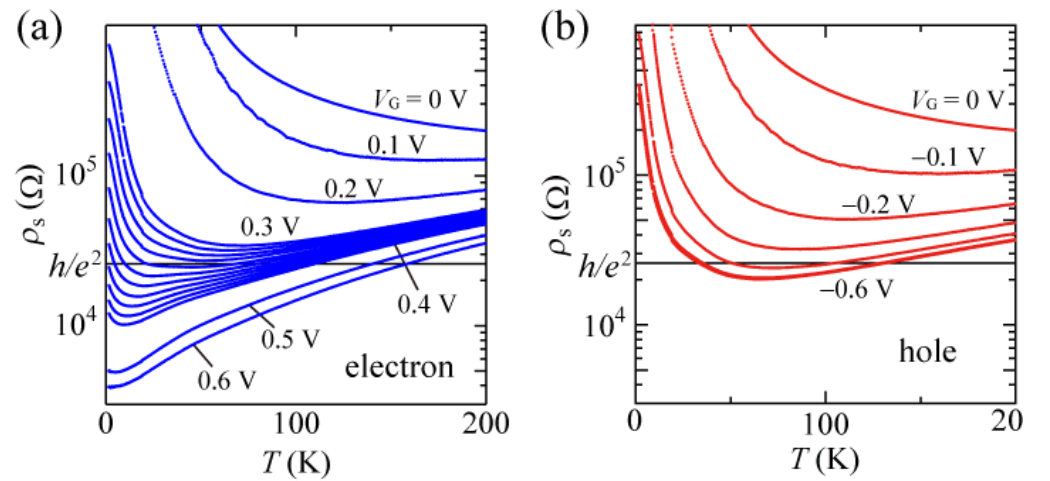

**Figure 9.** Temperature $T$ dependence of surface resistivity, $\rho_s$, under (**a**) electron doping and (**b**) hole doping.

### 3.2.2. Hall Coefficient

In the case of a Fermi liquid with a single type of carrier, $1/e|R_H|$ ($R_H$: Hall coefficient) would denote the carrier density corresponding to the volume enclosed by FS [24] and should be independent of temperature. On the contrary, temperature-dependent $R_H$, which was inconsistent with the volume of the noninteracting FS, often appeared in the normal state of strongly correlated materials. In the noninteracting single-band picture, κ-Cl had a large hole-like FS so that $1/eR_H$ in the metallic state should be $+p(1-\delta)$, where $p$ and $\delta$ are the half-filled hole density per layer and the electron doping concentration, respectively. Figure 10(b) shows the $V_G$ dependence of $1/eR_H$ at 40 K. Near the charge-neutrality point (Mott-insulating state), we could not observe the distinct Hall effect due to the high resistivity. Upon electron doping, the Hall coefficients became measurable and were positive despite the doped electrons. $1/eR_H$ appeared to obey $+p(1-\delta)$ under sufficient electron doping, indicating that the Mott-insulating state collapsed, and the system approached the metallic state based on the noninteracting band structure. However, upon hole doping, $1/eR_H$ became much less than $+p(1-\delta)$. The values also differed from the externally doped hole density, $-p\delta$, implying that the Mott-insulating state collapsed by hole doping, but the system approached a different electronic state with a smaller FS than the noninteracting case.

The temperature dependence of $R_{Hs}$ also revealed the peculiarity of the hole-doped state, as shown in Figure 10c. $R_{Hs}$ was almost temperature-independent under electron doping, as expected for a conventional metal. By contrast, $R_{Hs}$ under hole doping monotonically decreased with an increasing temperature, approaching values similar to those under electron doping.

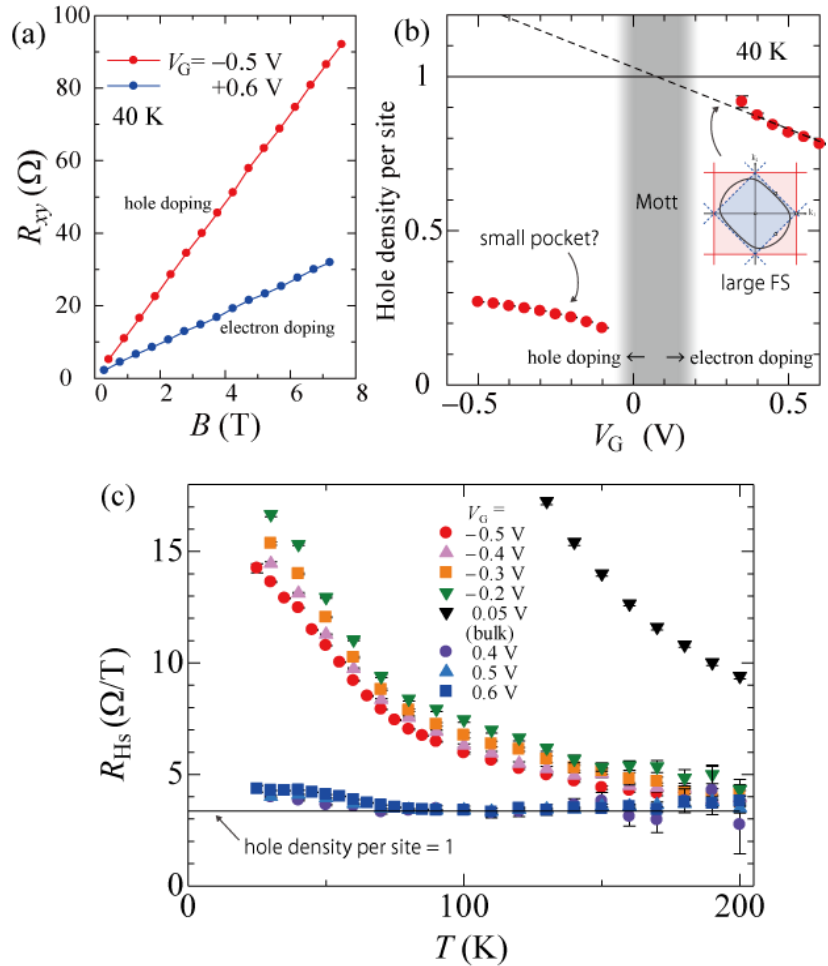

**Figure 10.** (**a**) Hall resistance vs. magnetic field at 40 K. (**b**) Gate-voltage dependence of the hole density per site (estimated from $1/eR_H$) at 40 K. The dashed line denotes the hole density per site estimated from the volume bounded by the noninteracting FS assuming that doping concentration is proportional to $V_G$ (20% doping at 0.5 V). The center of the shaded insulating region corresponds

to the charge neutrality point (resistivity peak). (**c**) Temperature dependence of $R_{Hs}$. The solid line indicates the value where the hole density per site becomes one (half filling).

3.2.3. Resistivity Anisotropy

In-plane conductivity anisotropy also reflected the anomalous state under hole doping. Figure 11 shows the in-plane anisotropy of the surface resistivity, $\rho_c/\rho_a$, up to 200 K. Here, $\rho_c$ ($\rho_a$) denotes the surface resistivity along the *c* axis [*a* axis; the short axis of the elliptical FS is parallel to the *c* axis, as shown in Figure 2b]. Under electron doping, the resistivity was almost isotropic ($\rho_c/\rho_a \sim 1$) and independent of temperature. Under hole doping, by contrast, $\rho_c/\rho_a$ was distinctly larger than one and increased with cooling. The conduction along the c-axis diminished in the hole-doped state, and its origin was weakened at high temperatures.

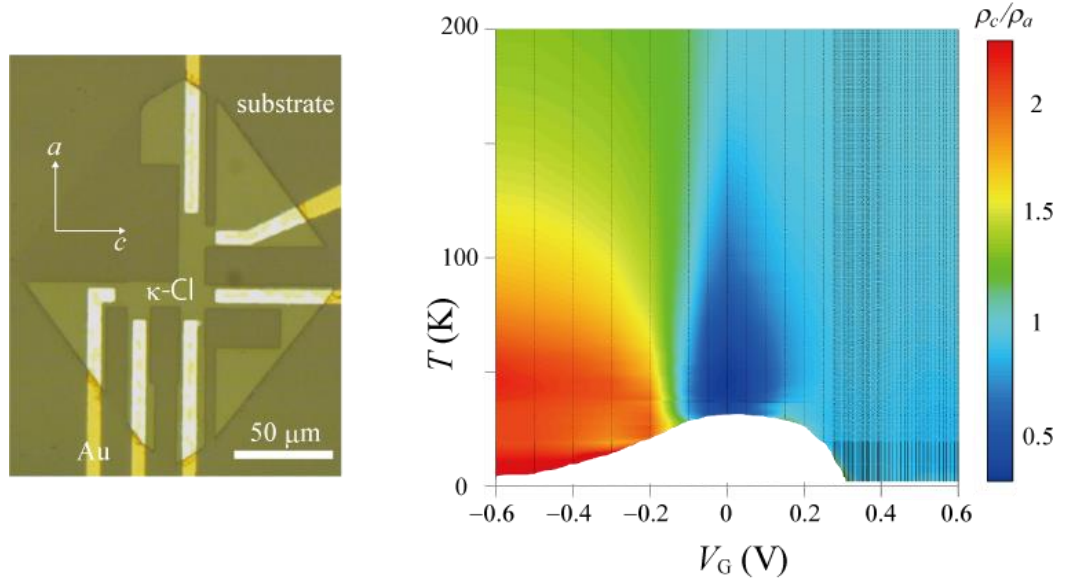

**Figure 11.** Temperature dependence of in-plane anisotropy of surface resistivity (note that both *a* and *c* axes are parallel to the conducting plane in this material). Data are missing at low temperatures and low doping (white region in the right panel) due to the high resistance.

3.2.4. Single-Particle Spectral Functions

We compared the above experimental results with model calculations. Figure 12 shows the single-particle spectral functions (corresponding to the DOS) of the Hubbard model on an anisotropic triangular lattice at 30 K using the cluster perturbation theory (CPT). The Mott-insulating state [Figure 12b,e] was reproduced at half filling, where the energy gap opened at all the k-points. When 17% of the electrons were doped [Figure 12c,f], the noninteracting-like FS emerged. On the other hand, the topology of FS under 17% hole doping [Figure 12a,d] appeared different from the noninteracting case. The spectral weight near the Z point was strongly suppressed (pseudogap), and a lens-like small hole pocket remained. The partial disappearance of FS was notable in the lightly hole-doped cuprates.

The calculated FS provides insights into the origin of the doping asymmetry of the Hall effect and the resistivity anisotropy. Sufficient electron doping reconstructed the noninteracting-like large hole FS, resulting in $1/eR_H \sim p(1-\delta)$. On the other hand, hole doping induced the partially suppressed lens-like small FS. In this state, Luttinger's theorem seemed violated, and $R_H$ could no longer be simply estimated. However, it was possible that $R_H$ was predominantly governed by quasiparticles with a relatively long lifetime (bright points of the spectral function in the reciprocal space in Figure 12), resulting in similar values of $1/eR_H$ corresponding to the area of the lens-like hole pocket. In addition, the large resistivity anisotropy under hole doping could be simply explained by the suppression of the quasi-one-dimensional FS along the Z–M line, which contributed to the conduction along the *c*-axis. As shown in Figure 12g, the spectral density on the Z–M line was recovered at high temperatures, consistent with the tendency in the Hall and anisotropy measurements.

The suppression of the spectral weight near the Z–M line under hole doping seemed to be related to the van Hove singularity (the van Hove critical points lie on the Z–M axis). With the doping of holes, FS approached the van Hove singularity, and the effect of the interaction was expected to be enhanced. It was also revealed that the spin fluctuation was stronger in the hole-doped state because of the van Hove singularity. By contrast, FS departed from the van Hove singularity with the doping of electrons, resulting in a weaker interaction effect and a more noninteracting-like FS.

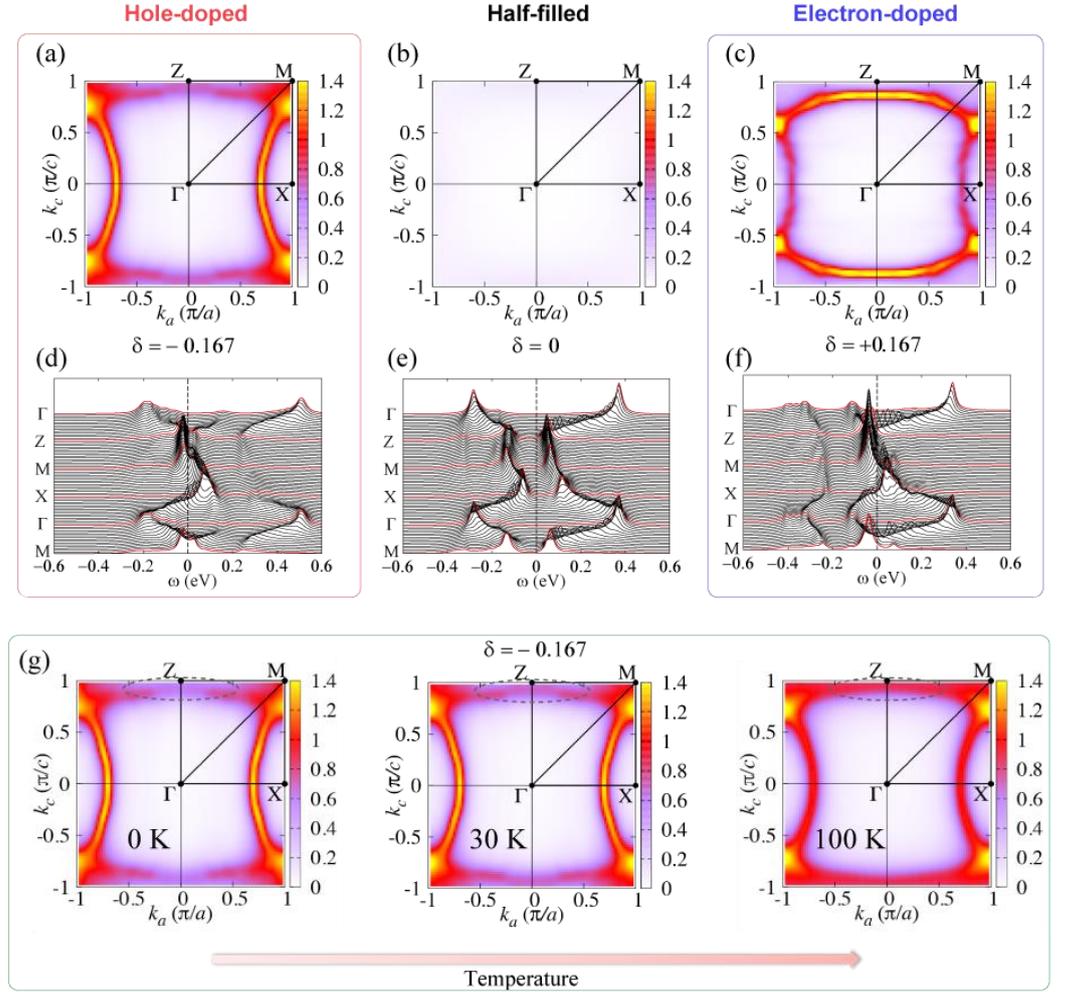

**Figure 12.** Fermi surfaces and single-particle spectral functions of the Hubbard model on an anisotropic triangular lattice at 30 K, calculated using the cluster perturbation theory (CPT). (**a**)–(**c**) Fermi surfaces for (**a**) 17% hole doping, (**b**) half filling, and (**c**) 17% electron doping, determined by the largest spectral intensity at the Fermi energy. (**d**)–(**f**) Single-particle spectral functions for (d) 17% hole doping, (**e**) half filling, and (**f**) 17% electron doping. The Fermi energy is located at $\omega = 0$ and the parameter set of this model is $t'/t = -0.44, U/t = 5.5$, and $t = 65$ meV. (**g**) Temperature evolution of the spectral density at 17% hole doping. The suppression near the Z–M line diminishes at high temperatures.

## 4. Summary

We fabricated bandfilling- and bandwidth-tunable EDL transistor devices using a flexible organic Mott insulator with a simple band structure. As shown in Section 3.1, drastic resistivity changes ranging from superconducting to highly insulating ($\rho > 10^9$ Ω) states occurred in the proximity of the tip of the Mott-insulating phase in the phase diagram. The superconducting phase surrounded the Mott-insulating phase in the band-filling–bandwidth phase diagram. The superconducting transition temperature, $T_C$, was almost identical among the electron-doped, hole-doped, and nondoped states, in contrast to those of the high-$T_C$ cuprates. Model calculations based on the anisotropic triangular lattice qualitatively reproduced the doping asymmetry on the doping levels for superconductivity and the tendency towards phase separation under electron doping, implying the significance of the flat part of the upper Hubbard band (originating from the noninteracting band structure). However, the calculations did not reproduce the reentrance into the slightly insulating state beyond the electron-doped superconducting state. One possibility was that a magnetic or charge-ordered state emerged at specific doping levels (for example, ~12.5%), as in the case of the stripe order in the cuprates [25–28].

At high $U/t$ where no superconductivity was observed, the transport properties exhibited non-Fermi liquid behaviors, as shown in Section 3.2. At high temperatures (above $T \sim 100$ K), the metallic-like conduction above the Mott–Ioffe–Regel limit (the bad-metal behavior) widely emerged regardless of the doping polarity, supporting the universality of the bad-metal behavior near the Mott transitions. At lower temperatures, the anomalously large, temperature-dependent Hall coefficient and in-plane resistivity anisotropy appeared under sufficient hole doping. Model calculations of the spectral density explained the anomaly under hole doping in terms of the partial disappearance of FS (pseudogap) due to the approach of the Fermi energy to the flat part of the energy band (same part as the flat part of the upper Hubbard band before doping). The pseudogap state also appeared in the lightly doped high-$T_C$ cuprates. However, the location of the pseudogap in k-space differed, owing to the difference of the noninteracting band structure, resulting in the different doping asymmetry.

The experimental methods shown here are applicable to other molecular conductors, including κ-(BEDT-TTF)$_2$Cu$_2$(CN)$_3$, which is a genuine Mott insulator without antiferromagnetic ordering [29]. The same experiments on this material may reveal more universal behaviors of a doped Mott insulator. Similar experiments on the molecular Dirac fermion system [30] are also possible and of great interest.

**Author Contributions:** Conceptualization, Y.K. and H.Y.; methodology, Y.K. and H.Y.; data curation, Y.K.; writing—original draft preparation, Y.K.; writing—review and editing, H.Y.; visualization, Y.K.; supervision, H.Y.; funding acquisition, Y.K. and H.Y. All authors have read and agreed to the published version of the manuscript.

**Funding:** This research was funded by MEXT and JSPS KAKENHI, Grant Numbers JP16H06346, JP19K03730, JP19H00891.

**Data Availability Statement:** All data needed to draw the conclusions in the paper are presented in the paper. Additional data related to this paper may be requested from the authors.

**Acknowledgments:** We would like to acknowledge R. Kato, K. Seki, S. Yunoki, J. Pu, T. Takenobu, S. Tajima, N. Tajima, and Y. Nishio for the collaborations and valuable discussions. The VCA and CPT calculations shown here were performed by K. Seki and S. Yunoki. This research was supported by MEXT and JSPS KAKENHI, Grant Numbers JP16H06346, JP19K03730, JP19H00891.

**Conflicts of Interest:** The authors declare no conflicts of interest.